# Enhanced Phonon Boundary Scattering at High Temperatures in Hierarchically Disordered Nanostructures

Dhritiman Chakraborty[*], L. de Sousa Oliveira, and Neophytos Neophytou

[1]School of Engineering, University of Warwick, Coventry, CV4 7AL, UK

[*] D.Chakraborty@warwick.ac.uk

## Abstract

Boundary scattering in hierarchically disordered nanomaterials is an effective way to reduce the thermal conductivity of thermoelectric materials and increase their performance. In this work we investigate thermal transport in silicon-based nanostructured materials in the presence of nanocrystallinity and nanopores at the range of 300 K – 900 K using a Monte Carlo simulation approach. The thermal conductivity in the presence of nanocrystallinity follows the same reduction trend as in the pristine material. We show, however, that the relative reduction is stronger with temperature in the presence of nanocrystallinity, a consequence of the wavevector-dependent ($q$-dependent) nature of phonon scattering on the domain boundaries. In particular, as the temperature is raised the proportion of large wavevector phonons increases. Since these phonons are more susceptible to boundary scattering, we show that compared to the wave-vector independent thermal conductivity value this $q$-dependent surface scattering could account for even a ~40% reduction in the thermal conductivity of nanocrystalline Si. Introduction of nanopores with randomized positions enhances this effect, which suggests that hierarchical nanostructuring is actually more effective at high temperatures than previously thought.

Keywords: thermal conductivity; phonon transport; boundary scattering; thermoelectrics; nanotechnology; nanocrystalline silicon; Monte Carlo simulations.

# I. Introduction

Hierarchical nanostructuring is the most promising approach to date to achieve very low thermal conductivities ($\kappa$) and high thermoelectric (TE) figure of merit (*ZT*) as hierarchically sized nanofeatures scatter phonons across wavelengths and reduce phonon transport throughout the spectrum [1, 2]. Materials with nanocrystalline boundaries, nanoinclusions, atomic defects, pores, voids, are common practice in new generation TE materials [3-7]. For a PbTe–SrTe system, for example this method has yielded a lattice thermal conductivity ($\kappa$) of 0.9 Wm$^{-1}$ K$^{-1}$ at 915 K and a *ZT* of 2.2 [2]. More recent studies, reported in the p-type Pb$_{0.98}$Na$_{0.02}$Te-SrTe system an extremely low $\kappa$ of 0.5 W K$^{-1}$m$^{-1}$ and a *ZT* of 2.5 at 923K [13], and a similarly low $\kappa$ = 0.55 W K$^{-1}$m$^{-1}$ in a SiGe nanoporous system [7]. Moreover, our prior works also point out that hierarchical nanostructures can improve the thermoelectric power factor as well [8-12].

From the theoretical point of view, a large number of works can be found in the literature in explaining thermal conductivity trends in nanostructured materials, by us and others, using a variety of theoretical and simulation methods [14-23]. In our recent works, for example, we provide a complete room temperature analysis of thermal conductivity trends in Si-based nanomaterials in the presence of nanocrystalline boundaries and nanopores [19, 22]. Here, we extend our prior Si-based studies to higher temperatures, and focus specifically on answering the following question: Are there any quantitative differences in the thermal conductivity trends in hierarchically nanostructured materials between what is observed at room temperature versus what can be observed at higher temperatures? In other words, are nanostructured features more effective, or less effective in reducing the thermal conductivity at higher, rather than lower temperatures? Our motivation resides on the following: (1) The average mean-free-path for phonon-phonon scattering at room temperature in Si is indicated in several reports to reside somewhere between 100 nm – 300 nm [24-28]. At elevated temperatures, the mean-free-path is reduced as the Umklapp 3-phonon scattering increases, increasing the importance of phonon-phonon scattering compared to phonon boundary scattering; but (2), on the other hand, as the temperature is raised, phonons of different frequencies and wavevectors (usually larger wavevectors) participate in transport, which react differently to scattering off the various nanofeatures, and specifically for this study, off nanocrystalline boundaries (usually increasing scattering probabilities). Therefore, in this work we focus on understanding how these two seemingly counter-acting mechanisms affect the thermal conductivity of hierarchically nanostructured materials at higher temperatures. Our results could help the design of better TE materials with ultra-low thermal conductivities, and the insight we provide could also be generalized for other materials, beyond Si, as well.

## II. Approach

For calculating the thermal conductivity we solve the phonon Boltzmann transport equation (BTE) using the Monte Carlo method, with all details and validation described in our recent works [19, 20, 22], therefore, below we provide the method details that are important in this work alone. The geometries we consider are shown in Fig. 1. For computational efficiency we consider a 2D simulation domain of length $L_x$ = 1000 nm and width $L_y$ = 500 nm, which is adequate to consider all the effects we are investigating. Fig. 1a shows a typical nanocrystalline geometry, whereas Fig. 1b shows a typical hierarchically nanostructured geometry, where nanopores of random position and diameter through a normal distribution are inserted.

In the nanocrystalline geometry cases, the average grain size in the simulation domain is defined as $<d> = L_x / <N_G>$, where $L_x$ is the length of the domain in the transport direction and $<N_G>$ is the average number of grains encountered in that length. Grains in the nanocrystalline case are generated using Voronoi tessellations, where grain boundaries are created by expanding grain areas radially outward from their initial "seeding points" until two areas meet [29], given initial input values for the number of "seeding points" and the dimensions of the domain. In these structures, the thermal conductivity is affected in two ways — the scattering of phonons at the grain boundary, and internal three-phonon Umklapp scattering inside the grains. Following the commonly employed boundary scattering picture, the scattering probability at the grain boundaries depends on the phonon wave vector, $q$, the roughness of the boundary, $\Delta_{rms}$, and the angle of incidence between the phonon path and the normal to the grain boundary, $\theta_{GB}$, as indicated in Fig. 1c. This determines whether an incident phonon will be transmitted to the other side or will be scattered, and is given by the commonly employed relation [29]:

$$t = \exp\left(-4\mathbf{q}^2 \Delta_{rms}^2 \sin^2\theta_{GB}\right) \quad (1)$$

If the phonon is scattered, another parameter which depends again on the roughness and the phonon wavevector, the specularity, $p$, determines the angle of phonon scattering [18]. $p$ takes values from 0 to 1, with $p = 0$ indicating a diffusive, randomized scattering angle. Thus, diffusely scattered phonons can also propagate along an angle other than the angle of incidence after scattering. $p = 1$ indicates specular reflection where the angle of incidence is the same as the angle of reflection. This is also applied to the boundaries of pores in the nanoporous cases (see Fig. 1b). Specifically, in the case of specular pore boundary scattering, the angle that the phonon will reflect into, $\theta_{ref} = 2\gamma - \theta_{inc}$, is defined based on geometrical considerations (the angle of incidence is the same as the angle of reflection) as:

$$\theta_{ref} = 2\gamma - \theta_{inc} \quad (2)$$

where $\theta_{inc}$ is the angle of propagation of the phonon relative to the x-axis, and $\gamma$ is the angle formed by the line perpendicular to the pore at the point of interaction and the x-axis (see Fig. 1d). In the case of the pores, however, we do not explicitly consider the $\mathbf{q}$-dependence, but we consider specular or diffusive boundary scattering just by assigning the specularity parameter $p$.

The main investigation we undertake in this work concerns the influence of boundary scattering at high temperatures, and more specifically the role that different wave vector ($\mathbf{q}$) and MFP phonons play in grain boundary scattering. Therefore, we find it useful to provide an indication of the participation of larger $\mathbf{q}$ phonons in thermal conductivity as the temperature is raised. We can show this for the pristine bulk material based on the lattice thermal conductivity, $\kappa_l$, as derived from the Boltzmann transport equation in the single mode relaxation time (SMRT) approximation. The phonon thermal conductivity is given as $\kappa_l = \sum_{i,q} \kappa'_{l,i}(\mathbf{q})$, where [30]:

$$\kappa'_{l,i}(\mathbf{q}) = k_B \tau_i(\mathbf{q}) v_{g,i}(\mathbf{q})^2 \left[\frac{\hbar\omega_i(\mathbf{q})}{k_B T}\right]^2 \frac{e^{\hbar\omega_i(\mathbf{q})/k_B T}}{\left(e^{\hbar\omega_i(\mathbf{q})/k_B T} - 1\right)^2} \quad (3)$$

In Eq. 3, above, the index $i$ refers to each mode or band (two transverse and one acoustic), $v_{g,i}(\mathbf{q}) = \partial \omega_i(\mathbf{q})/\partial \mathbf{q}$ is the group velocity as a function of the wavevector, and is derived from the expression provided for each band's $\omega$, $\tau_i$ is similarly the relaxation time for each mode, $T$ is the temperature, $k_B$ is the Boltzmann constant and $\hbar$ is the reduced Planck's constant. The Bose-Einstein statistics are incorporated in this equation in the so-called phonon window function term [30]:

$$W_p(\mathbf{q}) = \frac{3}{(\pi^2 k_B T)} \left[\frac{\hbar \omega(\mathbf{q})}{k_B T}\right]^2 \frac{e^{\hbar \omega(\mathbf{q})/k_B T}}{\left(e^{\hbar \omega(\mathbf{q})/k_B T} - 1\right)^2} \tag{4}$$

which is shown in Fig. 2a, and indicates the part of the phonon spectrum that takes part in transport as the temperature increases. Indeed, the proportion of large $\mathbf{q}$ phonons increases (which makes transmission over boundaries on average more difficult — see Eq. 1). On the other hand, since higher energy phonons are excited at higher temperatures, Umklapp scattering increases (decreasing scattering times), which reduces the phonon mean-free-paths (MFPs). To evaluate the combined effect of higher wavevector phonon participation and additional scattering to the lattice thermal conductivity with temperature, we computed $\kappa'_{l,i}$ as in Eq. 3. We adopted the phonon spectrum analytical expression $\omega(\mathbf{q}) = v_s \mathbf{q} + c\mathbf{q}^2$ with $v_s = 9.01 \times 10^3$ ms$^{-1}$ and c = -2 × 10$^{-7}$ m$^2$s$^{-1}$ for the longitudinal acoustic (LA) branch, and $v_s = 5.23 \times 10^3$ ms$^{-1}$ and c = -2.26 × 10$^{-7}$ m$^2$s$^{-1}$ for the transverse acoustic (TA) branches [20], as used within the MC simulator. For simple first-order demonstration, for the phonon Umklapp scattering relaxation time we adopted Holland's model [26, 30]:

$$\tau_U^{-1} = B \omega_i(\mathbf{q})^2 T \exp\left(-\frac{C}{T}\right) \tag{5}$$

where B = 2.8 × 10$^{-19}$ s/K and C = 140 K. (Note that coded in the Monte Carlo solver is a more detailed and elaborate scattering rate description as in [15, 18, 19]).

The relative contribution of each $\mathbf{q}$-state to the thermal conductivity as the temperature changes, compared to the 300 K contribution, is shown in Fig. 2b. To produce the plots in Fig. 2b, we first compute $\kappa'_{l,i}(\mathbf{q})$ at $T$ = 300, 400, 500, and 600 K using Eq. 4. Then we perform the following two steps: i) We normalize the $\kappa'_{l,i}(\mathbf{q})$ values with respect to their initial $q = 0$ value, $\kappa'_{l,i}(0)$, ii) We normalize again those new values with their corresponding values at 300 K (which we use as our basis). This tells us how much more higher $q$-states contribute to thermal transport at higher temperatures, compared to room temperature. Thus, Fig. 2b shows the redistribution of heat flow over the entire spectrum, indicating that high frequency phonon scattering has a greater overall contribution (up to 20 % for $T$ = 600 K) to thermal transport at higher temperatures. Note that although the Window function only captures Bose-Einstein statistics, the analysis in Fig. 2b also captures the increase in Umklapp scattering, which are competing processes in increasing/reducing the participation of high $q$-states. This makes boundary scattering more dominant in reducing the thermal conductivity at higher temperatures as we show below in the Monte Carlo simulations.

## III. Results

We begin our investigation with the effects of nanocrystallinity on the thermal conductivity of Si as the temperature is raised. In Fig. 3a, with the blue line we show the thermal conductivity $\kappa$ for the bulk pristine system (without nanocrystallinity) as a function of temperature from 300 K to 900 K, which has the typical reduction trend due to the increase in the phonon-phonon scattering rates. At 900 K, for example, there is a decrease in $\kappa$ by more than 60%.

The introduction of nanograin boundaries into the material (as shown in the inset of Fig. 3a) causes an additional reduction in $\kappa$ as expected. We further show in Fig. 3a the $\kappa$ as temperature increases from 300 K to 900 K for nanocrystalline (NC) geometries with <$d$> = 100 nm (Fig. 3a, green line) and <$d$> = 50 nm (Fig. 3a, red line). A sharp reduction in $\kappa$ of over 75% is observed at 300 K for the <$d$> = 100 nm case (green line) from the bulk value (blue line) at 300 K. The $\kappa$ drop is greater in the <$d$> = 50 nm case (red line) indicating as expected that a smaller grain size causes greater $\kappa$ reduction. Noticeably, however, at elevated temperatures, the reduction in $\kappa$ for the <$d$> = 100 nm NC structure (Fig. 3a, green line) drops by a larger amount (85% at 800K). This *enhanced* reduction of $\kappa$ observed at high temperatures cannot be explained based on phonon-phonon scattering and/or the reduction in <MFP> due to grain boundaries alone, but as we show further below, it is a consequence of the *q*-dependent boundary scattering, which increases as the average phonon *q*-value increases. This overall enhanced $\kappa$ reduction of ~ 7% of the pristine value (from 78% to 85% due to the *q*-dependent scattering processes) might seem small compared to just the pristine case values, however, in absolute terms that number is large if compared to the $\kappa$ of the NC case, which stands at just 16% of the pristine material at 300 K for $<d>$ = 100 nm. Essentially, this 7% reduction is almost half of the remaining $\kappa$ after the introduction of NC.

To clarify this magnitude, in Fig. 3b, we normalize the thermal conductivity to the 300 K value for both the single crystal and NC systems (i.e. at 300 K all geometries begin from the same reference), therefore, essentially *removing* the effect of the <MFP> reduction due to the grain boundary scattering. The difference in the blue line (pristine) versus the green/red lines (NC structures), is now the difference in the phonon-phonon scattering, and the *q*-dependent boundary scattering. By further normalizing $\kappa$ for the NC geometries by the pristine $\kappa$ value at all temperatures in Fig. 3c, i.e. by computing the $\kappa_{NC}/\kappa_{pristine}$ ratio at each temperature (where $\kappa_{pristine}$ is the blue line in Fig. 3b), we can further *eliminate* the effect of the temperature-dependent phonon-phonon scattering reduction which is common for all cases. Thus, the differences in the lines of Fig. 3c from the pristine blue-dashed line show the influence of the *q*-dependent boundary processes for the NC cases of $<d>$ = 100 nm (green line) and $<d>$ = 50 nm (red line). The *q*-dependent processes are responsible for doubling the reduction in $\kappa$ for the $<d>$ = 100 nm (green line) and amplifying the reduction in $<d>$ = 50 nm (red line) by a factor of four at 900 K. Note that this *q*-dependent scattering at the grain boundaries corresponds to a small decrease in $\kappa$ (compared to the pristine material). In the presence of crystalline boundaries where $\kappa$ is already reduced however, this is a large component. As a result, the *ZT* value of a thermoelectric material will benefit significantly due to this. Note that this effect is built in the boundary scattering, it is not something that can be easily used as a design strategy directly, but separating its effect from the geometrical features alone helps the understanding of phonon transport in nanostructures.

Next, we perform the same investigation for hierarchical nanostrucutred Si-based materials, in which case we incorporate nanopores into the nanocrystalline structures (as in Fig. 1b). Here we chose the pore diameter randomly from $D$ = 10 nm to $D$ = 50 nm using a uniform distribution. Again, we consider operating temperatures from $T$ = 300 K to 900 K. In Fig. 4a (blue line) we show the thermal conductivity versus temperature as in Fig. 3a, but for the new geometry (the blue line for the pristine material is the same in Fig. 3a). The green and red lines show the behavior of the hierarchical nanostructures (combined nanocrystalline and nanoporous (NC+NP) case) with $<d>$ = 100 nm (green-solid line) and $<d>$ = 50 nm (red-solid line). A reduction in $\kappa$ of over 80% is observed at 300 K for the $<d>$ = 100 nm case (green line), which is more than the NC only case. $\kappa$ drops further by more than 90% for temperatures at 800 K. The drop in $\kappa$ is even larger in the $<d>$ = 50 nm case (red line) at ~ 95%.

In Fig. 4b, as earlier in Fig. 3b, we normalize the thermal conductivity of the three structures to their value at $T$ = 300 K. Thus, we *remove* in this figure the effect of geometry and the MFP of boundary scattering due to the nanostructured features that the phonons encounter. What remains is then the effect of different phonon-phonon MFP (which we do not expect to have much difference) and the $q$-dependence of the boundary scattering. It is clear from this figure that as the temperature increases, the reduction in the thermal conductivity is larger in the nanocomposite structure, as also observed earlier. In dashed lines we also show the NC lines of Fig. 3b, indicating that the reduction is slightly larger in the hierarchically nanostructured materials.

This is more clearly shown in Fig. 4c, in which we normalize the data lines of Fig. 4b to the pristine blue line of Fig. 4b. This removes the influence of the phonon-phonon scattering, and what is left is just the influence that the $q$-dependence has on the boundary scattering as in Fig. 3c. Again, the $q$-dependence of the boundary scattering has a severe effect, even larger by ~15 % in the case of the hierarchical geometries (solid lines) compared to the nanocrystalline geometries (dashed lines). This is again expected to have a proportional contribution to *ZT*. We remark that while pores themselves do not have any $q$-dependent transmission properties (we treat pore scattering as $q$-independent), their presence introduces more disorder, increases the number of scattering events in the system and the times phonons pass through the grain boundaries, thus enhancing the overall effect of *T*-dependent surface scattering. Effectively, pores force phonons to pass from grain boundaries back and forth more times, which increases the importance of the $q$-dependent processes across the boundaries.

In order to directly compare and contrast $q$-dependent and $q$-independent scattering at grain boundaries, we turn off the $q$-dependence and re-run the simulations. In this case we consider only specular scattering at grain boundary and transmission without change in phonon angle. In this way, we compare the thermal conductivity of randomizing barriers versus q-dependent specular barriers, which also include temperature dependence. In Figs. 5a–c the effect of $q$-independent grain boundary scattering, for the nanocrystalline (NC) case with $<d>$ = 50 nm (purple-dashed line) is compared to the $q$-dependent case (red line). Figure 5a corresponds to same calculations performed for in Figs. 3b and 4b, now including the $q$-independent $<d>$ = 50 nm (purple line) calculation. Clearly, as expected, the $q$-independent specular simulation result resides between the pristine and the $q$-dependent randomized scattering result through the entire temperature range. Figure 5b is equivalent to Fig. 3c and Fig. 4c and shows the result of removing phonon-phonon scattering effects, indicating the difference between $q$-dependent and $q$-independent boundary scattering. This is shown more clearly in Fig. 5c, where we normalize the $q$-dependent system to the $q$-independent case (purple-dashed line) at 300 K. Note that this normalization that aligns the results at 300 K, also takes away the effect of the specular scattering being less severe compared to the randomized scattering on the thermal conductivity. What remains, is the increase in randomized scattering with temperature due the $q$-dependent specularity. In the nanocrystalline (NC) case that is $q$-dependent, there is a steady decrease in $\kappa$ as temperature increases, approaching a ~ 40% further reduction in comparison with the $q$-independent case. We also note that a similar treatment could have been done in simulations that compare a fixed transmission probability across a boundary, with $q$-dependent transmission. For fair comparison a justification of the fixed transmission value for all phonons would need to be provided, and thus we do not perform these simulations here. The $q$-dependent transmission, however, would make the effect we discuss even more pronounced.

## IV. Conclusions

In this work we employed the Monte Carlo transport simulator to solve the Boltzmann transport equation (BTE) for phonons in nanocrystalline Si-based materials and nanocrystalline materials with pores. We investigated the influence of wavevector $q$-dependence scattering on the nanocrystalline boundaries, and how this affects the thermal conductivity. We show that at higher temperatures, because the average phonon wavevector ($q$) increases, scattering throughout boundaries becomes stronger, and the thermal conductivity reduction is enhanced with temperature, compared to $q$-independent boundary scattering. We show that even up to ~ 40% further reduction in thermal conductivity at high temperatures (800 K) is attributed to the $q$-dependence of boundary scattering, compared to if the boundary scattering was $q$-independent. The introduction of random nanopores in addition to nanocrystallinity magnifies this effect by an additional ~15% at 800 K. This suggests that nanostructuring at high temperatures can actually be more effective than previously thought, and a simple constant MFP due to boundary scattering overestimates the thermal conductivity. We expect that our approach could be used to predict the behavior of materials depicting $q$-dependent boundary scattering at high temperatures, as well as providing insight into the design of low thermal conductivity, high *ZT* nanostructured materials.

# V. Acknowledgements

This work has received funding from the European Research Council (ERC) under the European Union's Horizon 2020 Research and Innovation Programme (Grant Agreement No. 678763). The authors would also like to thank Dr. Patrizio Graziosi at the University of Warwick for useful discussions.

## Appendix:

<u>Comparison of single-phonon Monte Carlo and literature data</u>

The figure below (Fig. A1) compares the single-phonon Monte Carlo thermal conductivity simulation results (blue line) for Si versus temperature to other (multi-phonon) Monte Carlo methods and experimental measured data. Overall very good agreement is found.

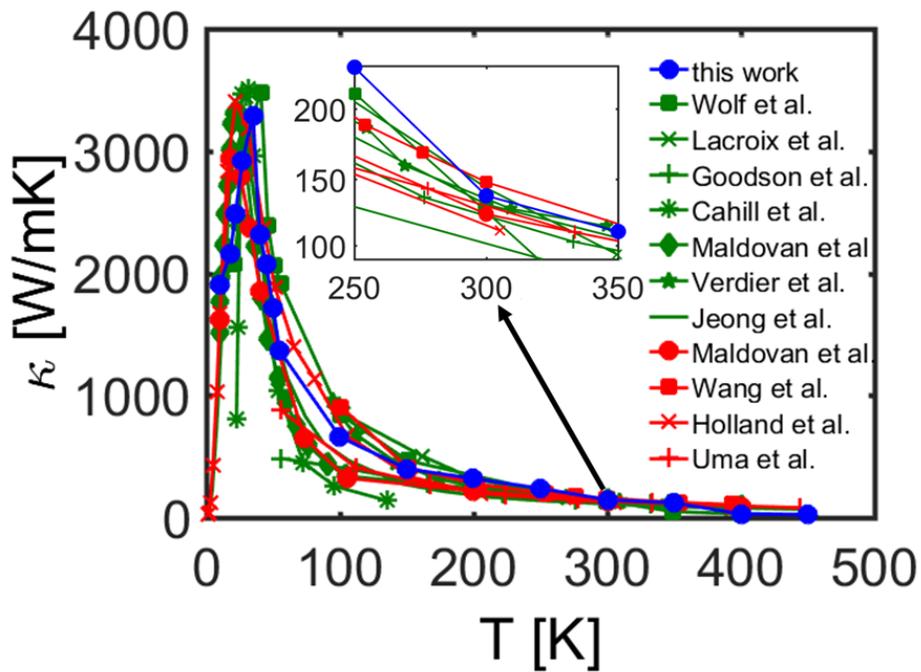

Figure A1. Validation of the single-phonon Monte Carlo simulator (blue line) for thermal conductivity ($\kappa$) versus temperature for bulk Si. Simulation works (green lines) [19, 15, 27, 31-35] and experimental works (red lines) [4, 26, 34, 36] are shown. The inset shows a close up of the temperature range between 250 K and 350 K. The simulated results are in close agreement to other literature results.

Figure 1:

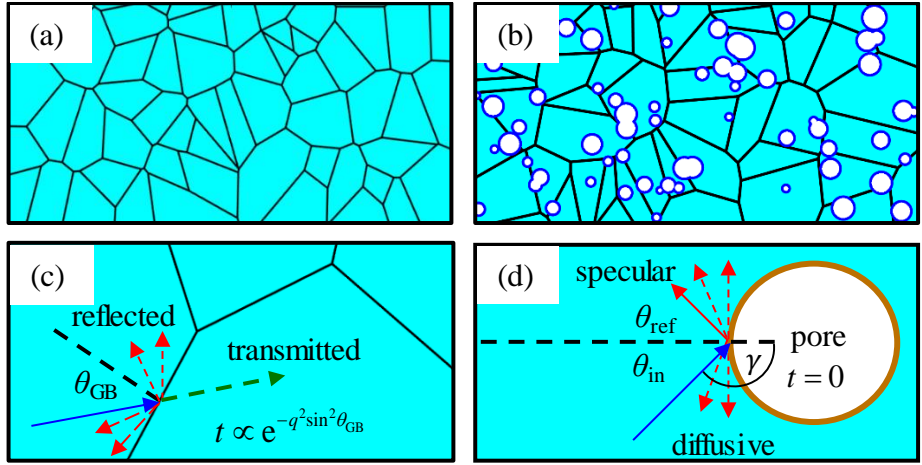

Figure 1 caption:

Examples of the nanostructured geometries considered. (a) Nanocrystalline (NC) materials, here grain dimension ($<d>$) is $<d>$ = 100 nm, black lines represent NC grain edges. (b) Nanocomposite material (NC+NP) with given grain dimension ($<d>$) and porosity ($\phi$). Here $\phi$ = 5%, for a random polydispersed pore arrangement, with pore diameter of uniformly distributed between 10 nm - 50 nm. (c) Schematic for grain scattering indicating the initial angle of the phonon, $\theta_{GB,}$ from the normal (dashed line), grain boundaries (black lines), initial path of the phonon (blue line) and probable paths of the phonon after scattering (red-dashed lines and green-dashed transmitted line). Transmission is dependent on grain boundary roughness as well as phonon wave vector $q$. Transmission probability is given by Eq 1. (d) Schematic for pore scattering indicating the pore boundary, the initial angle of the phonon $\theta_{in}$, and new angle of propagation after reflection, $\theta_{ref}$, depending on specularity parameter $p$. Probable paths of the phonon after scattering for both diffusive (red-dashed lines) and specular (red-solid line) are depicted. Transmission ($t$) through pore is always zero.

Figure 2:

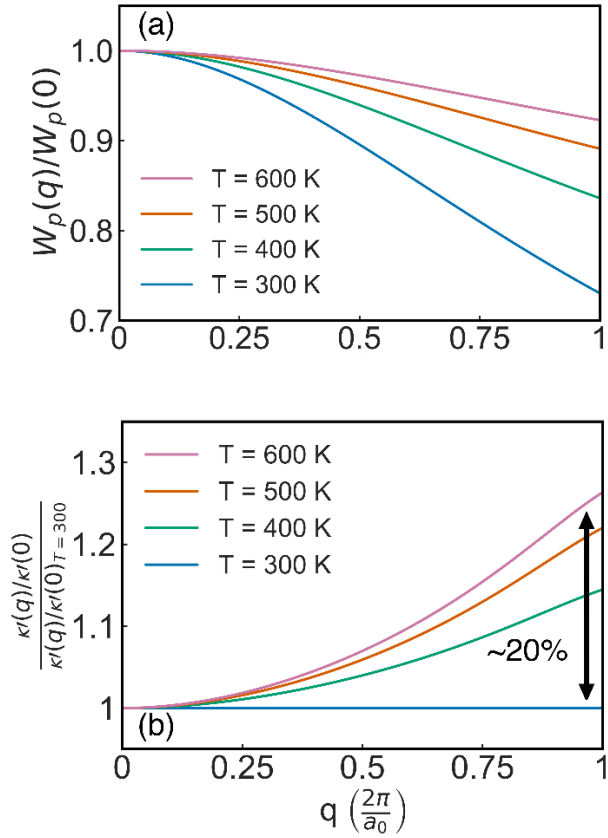

Figure 2 caption:

Change in average phonon attributes with increase in temperature. (a) Phonon window functions versus wavevector (normalized to unity) illustrating the range of phonon wavevectors active at different temperatures. As temperature increases the window function widens from the blue line at 300 K to the purple line at 600 K as the proportion of larger wavevector (shorter wavelength) phonons increases. (b) An indication of the redistribution of heat flow in the phonon spectrum, i.e. and indication of how much more higher $q$-states contribute to thermal transport at higher temperatures, compared to room temperature. We show normalized $q$-resolved thermal conductivity by its $\kappa'_{l,i}(0)$ value, and further normalized by the corresponding 300 K values. Temperatures $T = 400$ K (green line), $T = 500$ K (orange line) and $T = 600$ K (purple line) are shown, with respect to the 300 K values (blue line). Large wavevector phonons (higher $q$-phonons) have a greater contribution to thermal transport at higher temperatures compared to room temperature (up to 20 % for $T =$ 600 K).

Figure 3:

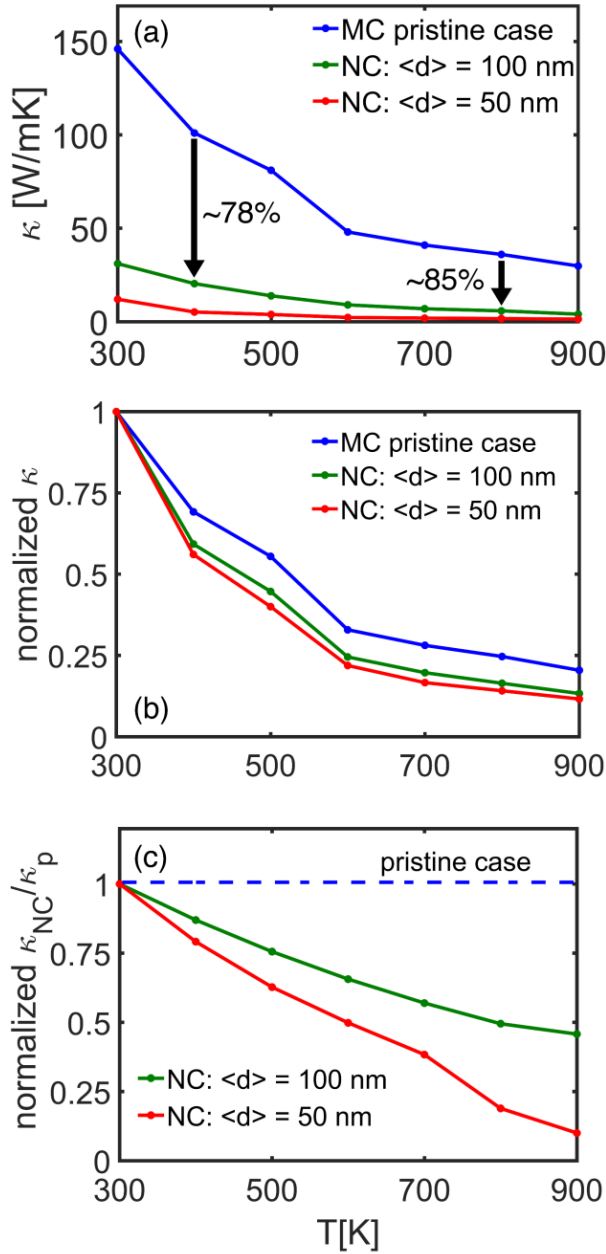

Figure 3 caption:

(a) Effect on thermal conductivity ($\kappa$) as temperature increases from 300 K to 900 K, for the pristine case (blue line), nanocrystalline (NC) case with $<d>$ = 100 nm (green line) and $<d>$ = 50 nm (red line). A sharp reduction in $\kappa$ of over 75% is observed at 300 K for the $<d>$ = 100 nm case (green line). This drops further to more than 85% for temperatures over 800 K. The $\kappa$ drop is greater in the $<d>$ = 50 nm case (red line). A typical geometry for $<d>$ = 100 nm case is given as Fig.1a. (b) The data in (a) normalized by the 300 K $\kappa$ value of the pristine case (blue line). The effect of phonon boundary scattering is removed from all data after this normalization. (c) The data in (a) normalized by the pristine value (blue line) at every temperature. The effect of phonon-phonon scattering from all data is taken away due to this normalization.

Figure 4:

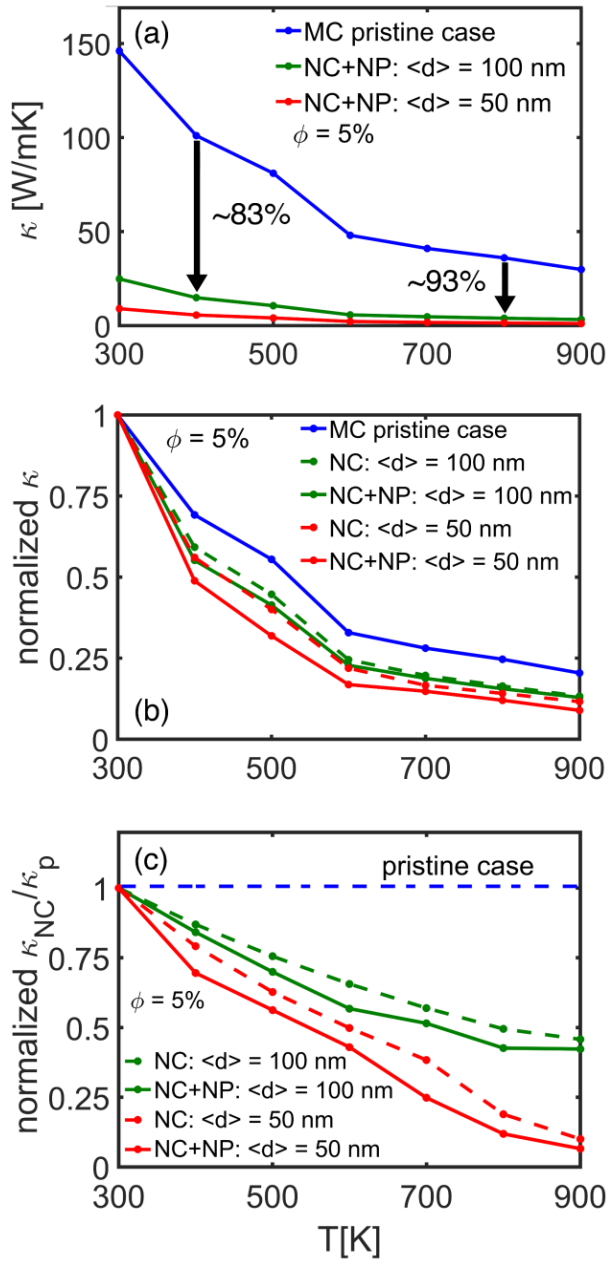

Figure 4 caption

(a) Effect of hierarchical nanostructuring on $\kappa$ as temperature increases from 300 K to 900 K, for the pristine case (blue line), combined nanocrystalline and nanoporous (NC+NP) case with $<d>$ = 100 nm (green-solid line) with porosity $\phi$ = 5 %, and $<d>$ = 50 nm (red-solid line) with porosity $\phi$ = 5 % . A reduction in $\kappa$ of over 80% is observed at 300 K for the $<d>$ = 100 nm case (green-solid line). $\kappa$ drops further by more than 90% for temperatures over 800 K. The $\kappa$ drop is greater in the $<d>$ = 50 nm case (red-solid line). A typical geometry for $<d>$ = 50 nm with $\phi$ = 5 % is shown in Fig.1b. (b) The data in (a) normalized by the 300 K $\kappa$ value of the pristine case (blue line). The effect of phonon boundary scattering is removed from all data after this normalization. (c) The data in (a) normalized by the pristine value (blue line) at every temperature. The effect of phonon-phonon scattering from all data is taken away due to this normalization. Again, the dashed lines are the data of Fig. 3(c) for the for the NC cases of $<d>$ = 100 nm (green-dashed line); $<d>$ = 50 nm (red-dashed line) alone. The legend of Fig. 4b applies one-to-one with the lines of Fig. 4c as well. At high temperatures there a further reduction in the normalized ratio observed due to pores.

Figure 5:

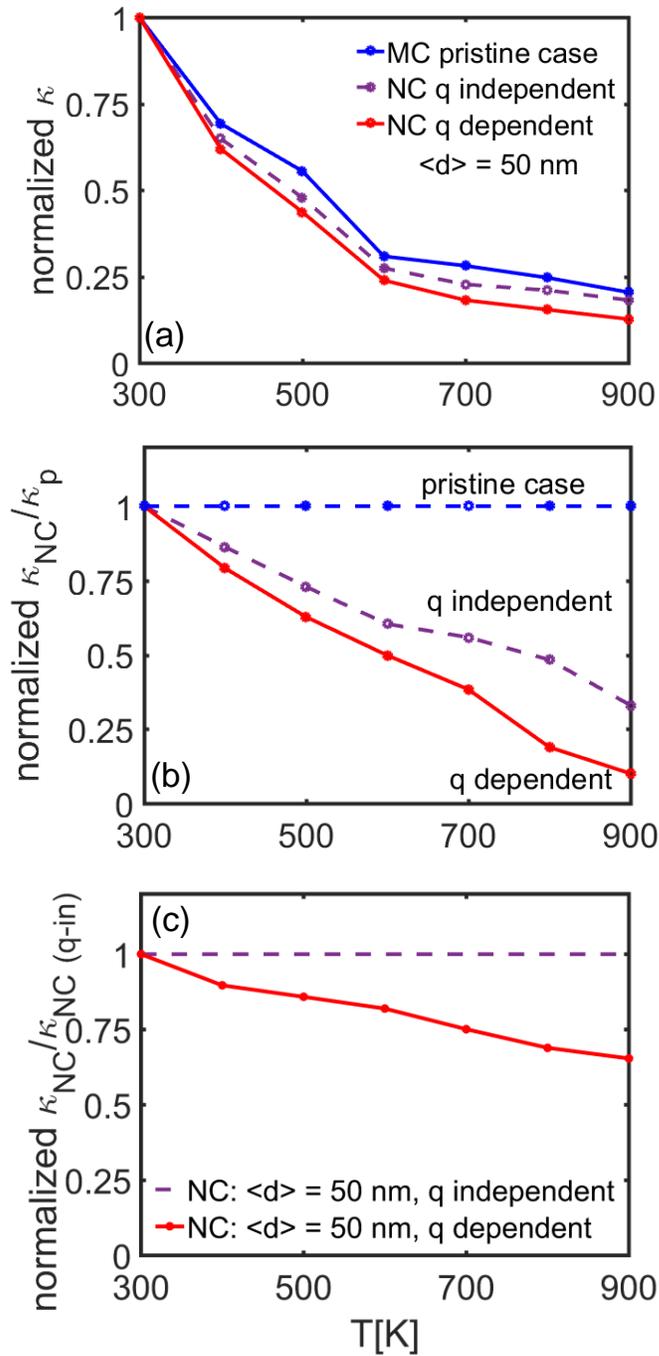

Figure 5 caption:

(a) The effect of $q$-independent and $q$-dependent scattering at grain boundaries compared to the pristine case. All values are normalized with respect to their value at 300K. The effect of phonon boundary scattering is removed from all data after this normalization. (b) The data in (a) normalized by the pristine value (blue line) at every temperature. The effect of phonon-phonon scattering from all data is taken away due to this normalization. (c) The data in (b) normalized to the $q$-independent NC case at all temperatures. This shows the effect of the $q$-dependence of the grain boundary scattering as the temperature increases.

\*\*\*